\documentclass{article}
\usepackage[utf8]{inputenc}
\usepackage{microtype}
\usepackage[margin=2.5cm]{geometry}
\usepackage{setspace}
\usepackage[none]{hyphenat}
\usepackage[pagewise]{lineno}
\usepackage[english]{babel}
\usepackage{color}
\usepackage{booktabs}
\usepackage{longtable}
\usepackage[most]{tcolorbox}
\usepackage{ragged2e}
\usepackage[overload]{empheq}
\usepackage[justification=justified]{caption}
\usepackage{enumitem}
\newtheorem{proposition}{Proposition}
\newtheorem{property}{Property}

\newtheorem{definition}{Definition}

\usepackage{amsmath,amsfonts,amssymb,mathtools}
\usepackage{cases}
\usepackage{multirow}
\usepackage{graphicx,float}
\usepackage{algpseudocode}
\usepackage[ruled,vlined, linesnumbered,nofillcomment]{algorithm2e}
\usepackage[hidelinks]{hyperref}
\usepackage{subcaption}

\newcommand{\cliset}{\mathcal N} 
\newcommand{\siteset}{\mathcal M}
\newcommand{\repset}{\mathcal R}
\newcommand{\s}[1]{\{#1\}}

\begin{document}
   
\begin{center}
  \begin{huge}
    {A rounding and clustering-based exact algorithm for the {p}-center problem}
\end{huge}
\end{center}

\begin{center}
    Zacharie {\sc Ales}$^{a,b}$,         Cristian {\sc Duran-Mateluna}$^{a,b,c,d}$,         {Sourour {\sc Elloumi}$^{a,b}$}
\vspace{.5cm}

\setstretch{0.8}{
{\it \relsize{-1}{$^a$: {UMA, ENSTA Paris, Institut Polytechnique de Paris, 91120 Palaiseau, France.}}\/}\\
{\it \relsize{-1}{$^b$: {CEDRIC,  Conservatoire National des Arts et M\'etiers, 75003 Paris, France.}}\/}\\
{\it \relsize{-1}{$^{c}$: {University of Santiago of Chile (USACH), Faculty of Engineering, Program for the Development of Sustainable Production Systems (PDSPS), Chile}}\/}\\
{\it \relsize{-1}{$^d$: {University of Santiago of Chile (USACH), Faculty of Engineering, Industrial Engineering Department, Chile}}\/}}
\end{center}

\rule{\linewidth}{0.1pt}
\noindent{\bf Abstract}

     The $p$-center problem consists in selecting $p$ facilities from a set of possible sites and allocating a set of clients to them in such a way that the maximum distance between a client and the facility to which it is allocated is minimized.   This paper proposes a new scalable exact solution algorithm based on client clustering and an iterative distance rounding procedure. {The client clustering enables to initialize and update a subset of clients for which the $p$-center problem is iteratively solved. The rounding drastically reduces the number of distinct distances considered at each iteration.} Our algorithm is tested on {396 benchmark instances with up to 1.9 million clients and facilities. We outperform the  two state-of-the-art exact methods considered when $p$ is not very small (i.e., $p>5$).}


\rule{\linewidth}{0.1pt}

\section{Introduction}

 The \emph{$p$-center problem} $(PCP)$ is a fundamental problem in  location science (\cite{laporte2019}). It consists of choosing  $p$ facility sites to minimize the maximum distance from a client to its nearest open facility. The $p$ open facilities are called $centers$.  Formally the problem is defined as follows. We consider a set of $N$ clients, a set of $M$ potential facility sites to open, and their corresponding index set  $\cliset=\{1,...,N\}$ and $\siteset=\{1,...,M\}$. Let $d_{i j}$ be the distance between client $i$ and site $j$, and $p$ be an integer. The objective is to find a set $S \subseteq \siteset$ such that $|S| \leq p $  and the value $z =\displaystyle\max_{i\in\cliset} \; (\displaystyle \min_{ j\in S} \; d_{ij})$  is minimized. This distance $z$ is called the $radius$ of solution $S$. 
 

The $p$-center is a NP-hard problem~(\cite{Kariv1079})   and is applied in various contexts, including facility location, emergency service planning, telecommunications network design, healthcare planning, and supply chain management. It allows to decide the best places for centers, such as warehouses or hospitals, to minimize distances or response times (\cite{ccalik2019p}).  There are a variety of exact and approximate methods for this classical problem and its variants. 



In this paper, we focus on the exact solution of large-scale instances. We combine existing ideas from the literature and introduce original ones in order to design an efficient solution algorithm.

Since the optimal radius is equal to one of the  distances $d_{ij}$, the first idea consists in considering a subset of the distances and deducing bounds from the result. This idea already appears in the binary search algorithm introduced in~\cite{Mineka1970} and~\cite{Daskin95a}, where for a chosen distance $\delta$ one can solve a set-covering problem to deduce whether the optimal radius is more than $\delta$ or not. This allows to iteratively divide the number of possible distances by two. This idea was for example extended by~\cite{Calik2013}  by only considering a subset of distances. We extend this idea by iteratively rounding down the distances with increasing precision, thus allowing us to drastically divide the number of distances at each iteration. 

{It is important to stress that for large-scale instances, the computation of $\s{d_{ij}}_{i\in \cliset,j\in \siteset}$ is prohibitive both in terms of time and memory.}
{To tackle this problem, the second idea that we leverage is  to only consider a subset of clients $\repset$, subsequently called \textit{representatives}. The optimal solution of the $(PCP)$ when only considering the clients $\repset$ enables to both obtain} a lower bound and an upper bound on the optimal radius. This idea has already been used in~\cite{Handler}  and~\cite{Chen2009} to start with an initial set of representatives and keep increasing it until the lower bound and the upper bound coincide. It was also recently used and improved  by~\cite{Contardo2019} to develop  one of the most effective methods of solving large-scale instances.
{Since computing all the distances is not possible for large instances}, we assume in the following that the instances are metric and that we are given as an input the coordinates of the clients and sites instead of their distances. This for example ensures that we are able  to compute distances between pairs of clients or the barycenter of a set of clients.

\vspace{0.8cm}
\noindent\textbf{{Contributions}}

{We introduce a new exact method for the $(PCP)$ which combines the two aforementioned ideas of considering a subset of distances and clients. Our main contributions are the following:}
\begin{enumerate}[itemsep=0.6pt]
    \item {We perform a more and more precise rounding of the distances which both enables to quickly provide strong bounds on the optimal solution value and to drastically reduce the number of distinct distances considered at any step.}

    \item {We use a clustering method to 
  initialize the subset $\repset$ of representatives of the clients. We additionally take advantage of the partition obtained by introducing the concept of cluster quadrants which enables us to update set $\repset$ by only adding a limited number of additional representatives.}

    \item {We design an efficient procedure to update the list of sites which are dominated. Depending on whether the representatives set $\repset$ is updated or the bounds on the optimal solution are improved, this strategy can significantly reduce the  number of sites to compare.}

    \item {We conduct an extensive computational analysis where we first assess the impact of different original features of our method. Then, we compare our algorithm to two state-of-the art methods~(\cite{Contardo2019,GAAR2022}), and we show that our method outperforms the other two methods when $p$ is not very small (i.e., greater than $5$). We consider substantially harder instances than those previously used to evaluate these two exact methods by increasing the maximal value of $p$ from $30$ to $200$ and the maximal number of clients and sites from one million to $1.9$ million, all methods being run on the same computer.}
\end{enumerate}



         
 The remainder of this article is organized as follows.  Section~\ref{sec:literaturePCP} reviews the related literature. Section~\ref{sec:cluteringPCP} presents our  exact solution method. Illustrations and extensive  computational results are presented in Sections~\ref{sec:illusPCP} and~\ref{sec:experPCP}. Finally, some conclusions  are drawn in Section~\ref{sec:conclusionPCP}.

\subsection*{Notations}

A solution of the $p$-center problem is represented by a vector $y\in\s{0, 1}^{M}$ such that $y_j$ is equal to~$1$ if and only if  site $j\in\siteset$ is opened in solution $y$. The set of opened sites in solution $y$ is denoted by $S(y)\subset\siteset$. For a given subset of clients $\repset\subseteq\cliset$, $rad_{\repset}(y)$ corresponds to the radius of $y$ when only the clients in $\repset$ are considered (i.e., $rad_{\repset}(y)=\displaystyle\max_{i\in \repset}\min_{j\in S(y)} d_{ij}$). Thus, the radius of solution $y$ is $rad_{\cliset}(y)$. The \textit{allocation distance} of a client $i\in\cliset$ in solution $y$ is  $\displaystyle\min_{j\in S(y)} d_{ij}$. A client is said to be \textit{covered within distance} $d\in\mathbb R$ by solution $y$ if its allocation distance is at most~$d$. Let $K$ be the number of distinct distances in $\s{d_{ij}}_{i\in\cliset,j\in\siteset}$ and let $\mathcal D=\{D^1,..., D^K\}$ be these distances  in increasing order.

For a given $\alpha\in\mathbb N$, {let $d^\alpha_{ij}$ be distance $d_{ij}$ rounded down to its nearest multiple of $10^\alpha$.} We denote by $(PCP_\alpha)$ the $p$-center problem in which distances $d^\alpha_{ij}$ are considered instead of $d_{ij}$. Let $rad^\alpha_{\repset}(y)$ be the radius of $y$ for clients $\repset$ in $(PCP_\alpha)$ (i.e., $rad^\alpha_{\repset}(y)=\displaystyle\max_{i\in \repset}\min_{j\in S(y)} d^{\alpha}_{ij}$).

\newpage
\section{Literature review}\label{sec:literaturePCP}


    
    Several MILP formulations of the $p$-center problem have been presented and discussed in the literature. The first and classical MILP formulation for the $(PCP)$ was proposed in \cite{Daskin} that consider integer variables to represent the location and allocation decisions, and one real variable to represent the radius. \cite{Elloumi2004} introduce an alternative formulation which relaxation bound dominates that of \cite{Daskin}. Other formulations are presented in~\cite{Calik2013},~\cite{ales2018compact}, and~\cite{GAAR2022}. We refer the reader to~\cite{duranmateluna:tel-04473412} for an extensive computational comparison of these formulations. The direct solution of any of these formulation in a standard solver fails at optimally solving instances with several thousands of clients and sites in reasonable time.


    A key idea to solve the $p$-center was introduced by~\cite{Mineka1970} and presented in~\cite{Daskin95a}. For a given distance $D^k\in\mathcal D$, determining if the optimal radius of the $(PCP)$ is at most equal to $D^k$ can be achieved through the solution of a minimum-cardinality set cover problem denoted by $(SC_k)$. In $(SC_k)$, one constraint is associated to each client $i\in\cliset$. This constraint imposes that at least one of the sites $\s{j\in\siteset~|~d_{ij}\leq D^k}$ must be opened. Consequently, if there exists a solution with at most $p$ open sites, then the optimal solution value of $(PCP)$ is  at most $D^k$. From this observation, \cite{Mineka1970} proposes to solve the $(PCP)$ by incrementing $k$ until $(SC_k)$ leads to a solution of value at most $p$.  As presented in Algorithm~\ref{alg:garf}, this can be improved by  solving the set cover problems through a binary search on  $\mathcal D$~(\cite{garfinkel1977}). To reduce the size of $\mathcal D$,   an initial solution is additionally computed  at Step~\ref{s:gub}.


        \cite{Daskin} also considered a  binary search approach, but solved maximum coverage problems instead of set cover problems.   \cite{ALK} propose implementation enhancements that reduce the number of iterations. \cite{Elloumi2004} improved  Algorithm~\ref{alg:garf} by first relaxing the integrity of the set cover variables in Step~\ref{s:sc}. This enables to quickly reach a very strong lower bound called $LB^*$. Different heuristics are also introduced to obtain an upper bound called $UB^*$. These bounds significantly reduce the possible values of the optimal radius index in $\mathcal D$ and thus the number of binary search iterations.

    \vspace{.2cm}
    
    \begin{algorithm}[H]
    \caption{\texttt{BinSearch}: solve $(PCP)$ by binary search~\cite{garfinkel1977}}
    \label{alg:garf}
    \DontPrintSemicolon
     \KwData{clients $\cliset$, sites $\siteset$, $p\in\mathbb N$}
    \KwResult{An optimal solution $y$ of $(PCP)$  and its radius}
        Compute $\s{d_{ij}}_{i\in\cliset,j\in\siteset}$\;
        Compute $\mathcal D=\s{D^1, ..., D^K}$ the sorted distinct distances in $\s{d_{ij}}_{i\in\cliset,j\in\siteset}$\;
        $y\leftarrow$ heuristic solution\label{s:gub}\;
        $\overline k\leftarrow $ distance index such that $rad_\cliset(y)=D^{\overline k}$\;
        $lb,ub\leftarrow 1,\overline k$\;
        \While{$lb<ub$}{
            $k\leftarrow \left\lfloor\frac{lb+ub} 2\right\rfloor$\label{s:setk}\;
    
            $\hat y, opt \leftarrow$ solve $(SC_k)$\label{s:sc}\;
            \eIf{$(SC_k)$ is feasible and $opt\leq p$}{
                $ub\leftarrow k$\;  
                $y\leftarrow \hat y$\;
            }{
                $lb\leftarrow k+1$\;
            }
        }    
      \Return $y, rad_\cliset(y)$\;
    \end{algorithm}
    
    \vspace{.3cm}

    In another approach, \cite{Handler} and \cite{Chen2009} presented iterative algorithms in which only a subset of clients $\repset$ is considered. Let $y_{\repset}$ be the optimal solution of the $(PCP)$ in which only clients $\repset$ are considered. They showed that $y_{\repset}$ is optimal for the original problem if and only if $rad_{\repset}(y)=rad_\cliset(y)$. Otherwise, they add a constant number of clients to  subset $\repset$.  
    
    Finally, to the best of our knowledge, the two  best exact methods also consider a progressively increasing subset of clients to solve the $(PCP)$. \cite{GAAR2022} explore a Benders decomposition of the classic formulation of \cite{Daskin}. They propose a method based on a specialized branch-and-cut that considers a lifting procedure for the Bender cuts. On the other hand, \cite{Contardo2019} leverage domination rules and a local search heuristic.  These two methods  can both solve instances with up to {one million} clients for small values of $p$ (i.e., $p\leq 30$).

    We now provide further details on the approach from~\cite{Contardo2019} as it shares similarities with the algorithm that we propose.

\newpage
\noindent\textbf{An existing row generation algorithm}
\label{sec:contardo}

The method proposed in~\cite{Contardo2019} is summarized in Algorithm~\ref{alg:contardo}. In this algorithm, the initial subset of clients $\repset$ is   obtained in Step~\ref{s:cinit} by randomly selecting $p+2$ clients, solving the $(PCP)$ on these clients and repeating the procedure until the radius of the worst solution obtained is not increased for a given number of iterations. The $p+2$ clients leading to the worst radius then constitute the initial set $\repset$.
 
\vspace{.4cm}

\begin{algorithm}[H]
\caption{Row generation algorithm from~\cite{Contardo2019}}
\label{alg:contardo}
\DontPrintSemicolon
\KwData{clients $\cliset$, sites $\siteset$, $p\in\mathbb N$}
\KwResult{An optimal solution $y$ of the $(PCP)$}

$\repset \leftarrow$  Initial subset of $\cliset$\label{s:cinit}\;
  $LB, UB \leftarrow  -\infty, +\infty$\;

  \While{$LB < UB$}{
    $\hat y, rad_{\repset}(\hat y) \leftarrow$ \texttt{BinSearch}($\repset$)\label{s:cpcp}\;
    $LB\leftarrow rad_{\repset}(\hat y)$\label{s:clb}\;

    \If{$rad_\cliset(\hat y)<UB$}{
        $y, UB\leftarrow \hat y, rad_\cliset(\hat y)$\label{s:cub}\;
    }
    
    \If{$LB < UB$}{ \tcc{Increase $\repset$}
      $\s{y^1, ...,  y^S} \leftarrow$  alternative  solutions obtained by  local search around $\hat y$\label{s:calt}\;
       Update $\repset$ by deducing from $\s{\hat y, y^1, ..., y^S}$ new clients to add \label{s:cupdate}\;
    }
  }
  \Return $y, rad_\cliset(y)$\;
\end{algorithm}

\vspace{.3cm}

At each iteration of the while loop, the $p$-center problem $(PCP_\repset)$ which only considers the clients in $\repset$ in solved (Step~\ref{s:cpcp}). To that end, the authors propose a variation of Algorithm~\ref{alg:garf}. In this variant, starting with $t=0$, the set covers $(SC_{2^t})$  are iteratively solved by incrementing $t$  until a solution of value at most $p$ is obtained. The last tested distance thus constitutes an upper bound on the optimal radius that enables to reduce the number of distance indices considered in the classical binary search algorithm. This provides a solution $\hat y\in\s{0, 1}^{\siteset}$. Since only a subset $\repset\subset \cliset$ is considered, the radius $rad_{\repset}(\hat y)$ constitutes a lower bound $LB$ on the optimal radius (Step~\ref{s:clb}). The radius $rad_N(\hat y)$, constitutes an   upper bound which may enable to decrease $UB$ (Step~\ref{s:cub}). 

To speed up the resolution of the set covers within the binary search, the authors also identified the non-dominated sites. 

\begin{definition} [\cite{church1984}]
    A site $j_1\in\siteset$ is said to be \textit{dominated} by another site $j_2\in\siteset$ relatively to a client subset  $\repset$ if 
     {$d_{i,j_1}\geq d_{i,j_2}~\forall i\in\repset$ and}
     either $\exists i\in\repset~d_{i,j_1}>d_{i,j_2}$ or $j_1<j_2$.
\label{def:domin}
\end{definition}

\noindent We can observe that with this definition,  two different sites cannot dominate each other. 
\begin{proposition} [\cite{church1984}]
There exists an optimal solution of $(PCP_{\mathcal R})$  in which no dominated site is opened.    
\end{proposition}


From this proposition, we deduce that solving  $(PCP_{\repset})$ for the non-dominated sites is the same as solving  $(PCP_{\repset})$ for sites $\siteset$.

If the optimality is not reached (i.e., $LB<UB$), the set of considered clients $\repset$ is increased. This step can have a significant impact on the solution time. Indeed, on the one hand, adding too much clients will slow down the solution of the $(PCP)$ at each subsequent iteration. On the other hand, adding to few or non-relevant clients may increase the number of remaining iterations required to reach an optimal solution. 

To add a relevant set of clients, the authors first  solve maximal coverage problems heuristically to generate a set of alternative solutions $Y=\s{y^1, ..., y^S}$. In each of these problems, the objective is to find a solution $y^h$ which enables to maximize the number of clients in $\cliset \setminus \repset$ covered within distance $LB$ (i.e., the optimal radius of $(PCP_\repset)$). Moreover, to ensure that $rad_{\repset}(y^h)=rad_{\repset},(\hat y)$ it is imposed in these problems that all clients in $\repset$ must be covered within distance $LB$ by $y^h$. Since the exact solution of such problems would be too time consuming, the authors solved them through a local search heuristic. 
In this heuristic, a  small set of clients  $C\subset\cliset\setminus \repset$ is first randomly selected. Solution $y^h$, which is initially equal to $\hat y$,  is then iteratively modified by:

\begin{enumerate}[itemsep=0pt]
\item randomly replacing a given number of sites in $y^h$; and then
\item iteratively applying the site replacement that maximizes the number of clients in $C$ covered within distance $LB$ until a local minima is reached.  
\end{enumerate}

If solution $y^h$ is the best currently known solution of the maximum coverage problem, it is added to $Y$. Moreover, if $y^h$ enables to cover all clients in set $C$ within distance $LB$, new clients are added to $C$. The algorithm stops when the size of $Y$  remains the same  for a given number of consecutive iterations. Finally, set $\repset$ is updated through a completion heuristic that identifies a subset of clients of $\cliset\setminus \repset$ that together would lead to an increase of the radius for all the solutions $\s{\hat y, y^1, ..., y^S}$ (Step~\ref{s:cupdate}).

We now present a new efficient exact algorithm based on a clustering of the client that takes also advantage of an iterative distance rounding procedure. 

\section{An exact rounding and clustering-based algorithm for (PCP)}\label{sec:cluteringPCP}

{This section is dedicated to the presentation of our exact solution method. A high-level view of it is described in Algorithm~\ref{alg:modulo}. We start by the clustering phase that partitions the set of clients into a given number $k\geq p$ of clusters. Then, the subset $\repset$ of representatives is initialized with the clusters medoids. The lower bound $LB$ is initialized $0$ and a feasible solution $y$ is chosen using $\repset$ to initialize the upper bound $UB$. These initialization steps~\ref{s:minit}--\ref{s:ubinit} are detailed in Subsection~\ref{sec:clustering}.}

{Then in the remaining steps~\ref{s:alpha}--\ref{s:decrement}, we perform our iterative rounding. The decreasing parameter $\alpha$ gives the precision of the rounding. It is initialized from the number of digits in $UB$. At each iteration, the crucial step is the exact solution of problem $(PCP_\alpha)$, i.e., the $p$-center problem with the rounding precision $10^\alpha$. This is detailed in Algorithm~\ref{alg:clustering} and in Subsection~\ref{sec:rounding}. At each iteration of this algorithm, the set of representatives $\repset$ is enlarged by adding, within each cluster, uncovered clients for different $(PCP)$ solutions. These solutions include the current solution as well as additional ones obtained by local search. We also introduce the notion of cluster quadrants in order to efficiently select the clients added to $\repset$. The update of the subset of clients $\repset$ is detailed in Section~\ref{sec:quadrant}.} {Another important aspect of our method is the elimination of dominated sites for the set of representatives (see Definition~\ref{def:domin}) within the solution procedure of each rounded problem $(PCP_\alpha)$. We put a focus on our novel ideas on this aspect in Subsection~\ref{sec:dom}.}

{We now present our new algorithm and its main features in Subsection~\ref{sec:clustering} to~\ref{sec:dom}.}


\subsection{The initialization steps}
\label{sec:clustering}

 {In Algorithm~\ref{alg:modulo}}, the initial subset of clients is not selected randomly{, as in Algorithm~\ref{alg:contardo},} but through the $k$-means algorithm~(\cite{macqueen1967some}) (Step~\ref{s:minit}). This clustering heuristic  provides a partition of the clients into $k$ clusters. As commonly done, we call the client that is the closest to the barycenter of each cluster its \textit{medoid}. The first advantage of using this clustering approach is that it enables to consider a subset of \textit{representative} clients $\repset$  of  $\cliset$ by initializing $\repset$ with the $k$-medoids of the partition (Step~\ref{s:minit2}). The second advantage is that each non-representative client in $\cliset\backslash\repset$ is now allocated to a cluster which we will leverage to update $\repset$. The number of clusters $k$ considered is a parameter of our method. Similarly to~\cite{Contardo2019}, we observed empirically that the value $p+2$ leads to good results in average.

\begin{algorithm}[H]\caption{Function   \texttt{solveByRounding}  that
optimally solves the $(PCP)$ with an increasingly precise rounding of the distances}\label{alg:modulo}
\DontPrintSemicolon
\KwData{clients $\cliset$, sites $\siteset$, $p$, $k \geq p$}
\KwResult{An optimal solution $y$ of the $(PCP)$}

  Clustering: use the $k$-means heuristic to obtain a partition $\mathcal C$ of $\cliset$ into $k$
  clusters\label{s:minit}\; 
  $\repset \leftarrow$ the $k$ medoids of clustering  $\mathcal C$ \label{s:minit2}\;   
  $y \leftarrow$ build a feasible solution from $\repset$ \; 
  $LB, UB \leftarrow 0, ~ rad_{\cliset}(y)$ \label{s:ubinit}\;

  $\alpha\leftarrow$ (number of digits  in $UB$)~ - 1  {// Used to round the distances following Definition~\ref{def:distanceRounding}}\label{s:alpha}\;
  
  \While{$ LB < UB$}{
    $ \hat y,rad_{\cliset}^\alpha(\hat y) \leftarrow  $\texttt{solvePCP}$_\alpha$($\cliset, \siteset, p, \repset, \mathcal C, \alpha$)  by Algorithm~\ref{alg:clustering}\;
    $LB \leftarrow rad_{\cliset}^\alpha(\hat y)$ \label{s:lb}\;
    \If{$rad_\cliset(\hat y) < UB$ \label{s:if}}{
    $y, UB \leftarrow \hat y, rad_\cliset(\hat y)$\label{s:newUB}\;
    }
    $\alpha\leftarrow \alpha - 1$ \label{s:decrement}\;
  }
  \Return Solution $y$

        \end{algorithm}

\vspace{.2cm}

 \subsection{{Exact solution of the \texorpdfstring{$p$}{}-center problem for the rounding precision \texorpdfstring{$10^\alpha$}{}}}\label{sec:rounding}

The bottleneck of  Algorithm~\ref{alg:modulo} is the solution of the $(PCP)$ subproblems through binary search (Step~\ref{s:cpcp}). Since the number of binary search iteration is in $\mathcal O(log (|\mathcal D|))$, we reduce the size of $\mathcal D$ by iteratively considering an increasingly precise rounding of the distances. 

More precisely, for a given rounding factor $\alpha\in\mathbb N$, each distance $d_{ij}$ is rounded down to the nearest multiple of $10^\alpha$.

\begin{definition}[{Distance rounding}]
    For a given client $i\in\cliset$ and a given  site $j\in\siteset$, {together with a lower bound $LB$ and an upper bound $UB$ on the optimal value of $(PCP)$}\[d^\alpha_{ij} \triangleq \min \left(\max \left( LB, 10^\alpha \left\lfloor\frac {d_{ij}}{10^\alpha}\right\rfloor \right), ~UB+1\right).\]
\label{def:distanceRounding}
\end{definition}
\vspace{-1cm}

The distance $d^\alpha_{ij}$ {is imposed to be} in the interval $[LB, UB+1]$ in order to further reduce $|\mathcal D|$.

The rounding factor $\alpha$ is initially equal to  the number of digits in the radius of a heuristic solution $y$ minus one (Step~\ref{s:alpha}). It is then decremented at each iteration thus enabling the $(PCP)$ to be solved with a more precise rounding at the next iteration (Step~\ref{s:decrement}).

At Step~\ref{s:clusteringAlpha}, we use function \texttt{solvePCP}$_\alpha$ to compute an optimal solution of the $(PCP)$ problem $(PCP_\alpha)$ in which each distance $d_{ij}$ is replaced by $d^\alpha_{ij}$. Note that this function could be replaced by any $(PCP)$ solution method (e.g.~Algorithm~\ref{alg:contardo}) provided that the distances can be rounded down whenever they are computed. Function \texttt{solvePCP}$_\alpha$ takes as an input both clients sets $\repset$ and $\cliset$. It considers $\repset$ as the initial set of clients and update it by adding clients from $\cliset \setminus \repset$  until an optimal solution $\hat y$ of $(PCP_\alpha)$ is obtained.

The following property ensures that optimally solving $(PCP_\alpha)$ enables to obtain a lower bound on the optimal radius of the original problem~(Step~\ref{s:lb}).

\begin{property}
    For any solution $\hat y$, $rad_\cliset^\alpha(\hat y)\leq rad_\cliset(\hat y)$.
    \label{prop:solalpha}
\end{property}

\noindent \textbf{Proof.}  Since $d^\alpha_{(ij)}\leq d_{ij}$ for all $ij\in\cliset\times\siteset$, then $\displaystyle rad_\cliset^\alpha=\max_{i\in\cliset}\min_{j\in S(\hat y)} d^\alpha_{ij}\leq \max_{i\in\cliset}\min_{j\in S(\hat y)} d_{ij}=rad_\cliset(\hat y)$. 

\hfill$\square$

After obtaining the lower bound $rad^\alpha_\cliset(\hat y)$, we 
compute the radius of  solution $\hat y$ without rounding down the distances (Step~\ref{s:if}) which constitutes an upper bound. If it is currently best known, we update $UB$ and $\hat y$ (Step~\ref{s:newUB}). Note that since the radius of $\hat y$ on the rounded down distances is equal to $LB$, the new value of $UB$ is at most $LB+10^\alpha$. Thus, throughout its iterations, this algorithm provides an increasingly tight guarantee on the quality of its bounds.  Moreover, at each iteration, there can only be at most $10$ distinct distances $d^\alpha_{ij}$.

Note that we only cluster the clients once at Step~\ref{s:minit} of Algorithm~\ref{alg:modulo}. Since the clients added to $\repset$ within function \texttt{solvePCP}$_\alpha$ are likely to be relevant for the subsequent iterations, they remain in $\repset$ until the end of the algorithm.

We now present function \texttt{solvePCP}$_\alpha$, detailed in Algorithm~\ref{alg:clustering},  that  takes advantage of the initial client clustering to update the subset of clients~$\repset$. This function starts by computing for each client $i\in \repset$ and each site $j\in\siteset$ the distance $d^\alpha_{ij}$.

\vspace{0.5cm}
\noindent\textbf{(PCP) relaxation}

{In Step~\ref{s:enti} of Algorithm~\ref{alg:clustering}, an optimal solution $\hat y$ of an integer $(PCP_\alpha)$ for clients $\repset$ is computed. Since this is the most time consuming step of our algorithm, we instead solve the linear relaxation of $(PCP_\alpha)$ whenever possible. The integer $(PCP_\alpha)$ is only considered if the linear relaxation solution $y^r$ does not enable to deduce any new client to add to $\repset$.}

\begin{algorithm}[H]\caption{Function \texttt{solvePCP}$_\alpha$ that optimally solves $(PCP_\alpha)$.}\label{alg:clustering}
\DontPrintSemicolon
                \KwData{clients $\cliset$, sites $\siteset$, $p\in\mathbb N^*$, $\repset\subset \cliset$, $LB\in\mathbb N$, $UB\in\mathbb N$, $\alpha\in\mathbb N $}
\KwResult{An optimal solution $y$ of $(PCP_\alpha)$}

  Compute $\s{d^\alpha_{ij}}_{i\in \repset,  j\in \cliset}$\label{s:d2}\;

  $J^\alpha_\repset \leftarrow$  non-dominated sites for $\repset$\label{s:dom}\;

  $addedClients \leftarrow  true$\label{s:irinit}\;

$y\leftarrow \emptyset$\;

  \While{$LB < UB$}{

 \eIf{addedClients}{
    $LB^r, y^r \leftarrow$  solve the relaxed $(PCP_\alpha)$ for $\repset$ and $J^\alpha_\repset$ by relaxed binary search\label{s:clusteringAlpha}\;
    $LB\leftarrow LB^r$\;
    $\hat y\leftarrow$ compute from $y^r$ an integer solution\label{s:conti}\;
  }{
    $\hat y, rad_{\repset}^\alpha(\hat y)\leftarrow$  solve the integer $(PCP_\alpha)$ for $\repset$ and $J^\alpha_\repset$ by binary search\label{s:enti}\;
    $LB\leftarrow rad_{\repset}^\alpha(\hat y))$\;
  }

  \If{$rad_{\cliset}^\alpha(\hat y)<UB$}{
    $y, UB\leftarrow \hat y, rad_{\cliset}^\alpha(\hat y)$\;
  }
  
  \If{$LB < UB$}{
    $\s{y^1, ..., y^S} \leftarrow$ alternative solutions obtained by local
    search around $\hat y$\label{s:alt}\;

    $addedClients \leftarrow$ Update $\repset$ by adding at most one client for each quadrant of each cluster and each solution $\s{\hat y, y^1, ..., y^S}$\label{s:quadrant}\;
  
    \If{$LB$ or $UB$ has been improved}{
      Clip the distances in the interval~$[LB, UB+1]$ and update the non dominated sites $J^\alpha_\repset$\;
    }
  }
  }
  \Return $y, rad^\alpha_\cliset(y)$\;
        \end{algorithm}

\newpage

To that end, we define a boolean parameter $addedClients$ equal to true if and only if clients have been added to $\repset$ at the last iteration. The variable $addedClients$ is initially set to $true$ (Step~\ref{s:irinit}). Thus, in the first iteration the linear relaxation of $(PCP)$ for clients $\repset$ and sites $J_\repset$ is solved (Step~\ref{s:conti}). This variable is then set to $false$ only if no client is added to $\repset$ at Step~\ref{s:quadrant}. In that case, the integer  $(PCP_\alpha)$ is solved at the next iteration (Step~\ref{s:enti}).

At Step~\ref{s:alt} of Algorithm~\ref{alg:clustering}, we use the same local search approach as in Algorithm~\ref{alg:contardo} from~\cite{Contardo2019}
to deduce from an integer solution $\hat y$, a set of alternative solutions $\s{y^1, ..., y^S}$. However, when we solve the linear relaxation of a $(PCP)$ at Step~\ref{s:conti}, we compute a fraction  solution $y^r$. To build an integer solution $y$ from $y^r$, we open the $p$ sites which values are maximal in~$y^r$ (Step~\ref{s:conti}). 

{We observed empirically that the value $LB$ obtained when solving $(PCP_\alpha)$ at Steps~\ref{s:conti} or~\ref{s:enti} is often the same for two consecutive iterations. Indeed, adding clients often does not lead to an increase of the optimal radius. Consequently, {at the first iteration of the binary search Algorithm~\ref{alg:garf}, we set $k$ to the value $1$ instead of  $\left\lfloor\frac{lb+ub} 2\right\rfloor$ (Step~\ref{s:setk}). As a consequence, $D^k$ is equal to $LB$ and if the solution of the set cover at Step~\ref{s:sc} does not require} more than  $p$ sites, we directly know that the optimal radius of $(PCP_\alpha)$ is not increased since the last iteration. This proved to be more efficient than the variation of the binary search considered in~\cite{Contardo2019} and described in Section~\ref{sec:contardo}.}

\subsection{Updating set \texorpdfstring{$\repset$}{} through the clusters quadrants and local search}
\label{sec:quadrant}
The choice of the clients added to $\repset$ at each iteration has a crucial impact on our algorithm. Indeed, increasing $\repset$ will tend to slow down the solution of the $(PCP)$ which would be an incentive not to add too many clients. However, adding too few clients will make the lower bound $LB$ increase more slowly, thus increasing the number of iterations. 

To find a relevant compromise, we can first observe that only adding to $\repset$ clients which are covered within $LB$ in the current solution $y$ cannot lead to an improvement of $LB$ at the next iteration (since the same solution $y$ would still lead to  a radius of value $LB$). Thus, we focus on the clients of $\cliset\setminus \repset$ which are not covered in $y$ within distance $LB$. Let $\cliset'(y)$ be this set of clients.

\begin{figure}[htpb]
\centering\begin{subfigure}[b]{0.8\textwidth}
    \centering\includegraphics{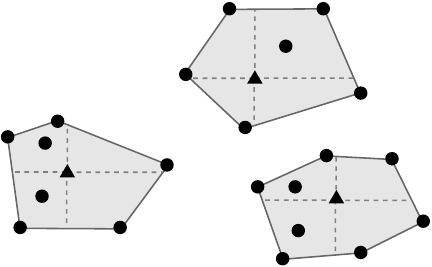}
    \caption{Three clusters and their medoids ($\blacktriangle$) which are included in $\repset$. The other clients in the cluster ($\bullet$) are not included in $\repset$. The clusters are represented by the convex hull of their clients and their quadrants are delimited by dotted lines.}
    \label{fig:quadrants_a}
\end{subfigure}
\centering\begin{subfigure}[b]{0.8\textwidth}
    \centering\includegraphics{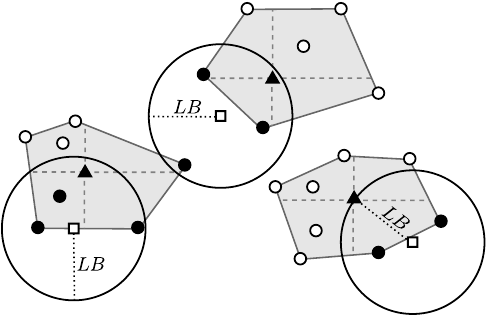}
    \caption{The squares represent the opened sites in solution $y$. The white circles correspond to the clients $\cliset'(y)$ which are not covered within distance $LB$ in solution $y$.}
    \label{fig:quadrants_b}
\end{subfigure}
\centering\begin{subfigure}[b]{0.8\textwidth}
    \centering\includegraphics{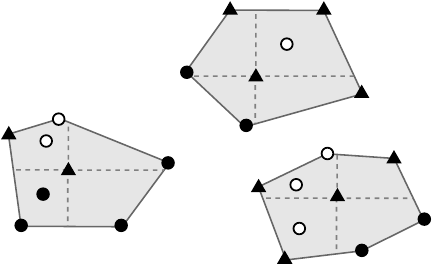}
    \caption{The  three clusters after their update. Set $\repset$ now contains $10$ clients.}
    \label{fig:quadrants_c}
\end{subfigure}
    \caption{Example of an update of a set $\repset$ and three clusters. }
    \label{fig:quadrants}
\end{figure}

Since the  $k$-means algorithm provides a coherent partition of the clients, we independently search  in each cluster for clients to add to set $\repset$. 
 For each cluster, we want to add clients which are as far away from their medoid as possible (so that they are more likely to increase the radius) and which are also as far away from each other as possible (to make it harder to cover them without covering other points in the cluster). Consequently, as represented in Figure~\ref{fig:quadrants_a}, we divide the clients of each cluster in four \textit{quadrants} according to their position in relation to that of their medoid. Figure~\ref{fig:quadrants_b} represents the solution $y$ in which three sites are opened. The clients represented by a white circle correspond to the set $\cliset'(y)$. We can see in the left cluster that  a client can be covered within distance $LB$ by a different site than the medoid of its cluster. For each quadrant, we add to $\repset$ the client of $\cliset'(y)$ which is the furthest away from the representative of the cluster, if any. For example, in Figure~\ref{fig:quadrants_c}, three clients are removed from two of the clusters and only one client is removed for the leftmost cluster. We apply this process to all solutions  $\hat y, y^1, ..., y^S$ (Step~\ref{s:quadrant}).

\subsection{Improving the computation of the sites dominations}
\label{sec:dom}
In Algorithm~\ref{alg:clustering}, whenever the lower bound $LB$ or the upper bound $UB$ is  improved, each distance  $d^\alpha_{ij}$ is updated according to Definition~\ref{def:distanceRounding}. This enables to both reduce the number of distances considered in the binary search algorithm and potentially increase the number of dominated sites.

 To reduce the computation time, we first observe that between two consecutive iterations, the set of dominated sites $J_\repset$ only needs to be updated if either:
\begin{enumerate}
    \item one of the bounds $LB$ and/or $UB$ has been improved;
    \item the number of considered clients $\repset$ has been increased.
\end{enumerate}


Moreover, in both cases, we do not need to test the domination of all possible couples of sites. More precisely, when the first case occurs, we only need to check if the non-dominated sites become dominated. Indeed, improving the bounds reduce the number of distinct distances, thus a dominated site remains dominated. Furthermore, we do not need to check if a non-dominated site is dominated by a dominated site.

In the second case, we only need to check if dominated sites become non-dominated. Indeed, adding clients to $\repset$ cannot make a non-dominated site dominated\footnote{Except in the case where two sites had the same distance to all clients before $\repset$ was updated.}. To improve the computation of the dominances in that case, we define the vector $dom\in\mathbb N^{M}$ such that $dom_j$ is equal to $0$ if site $j$ is not dominated and to the lowest index of a site that dominates $j$ otherwise. If $dom_j\neq 0$, we can avoid testing if site $j$ is dominated by sites $\s{1, ..., dom_j-1}$. Indeed, if all these sites were not dominating $j$ before the addition of the clients, they can not dominate it after either.

\newpage 
\section{An illustrative example} \label{sec:illusPCP}

{We illustrate our algorithm by considering the TSPLib~\cite{TSPlib} instance \texttt{nu3496} which contains the coordinates of $3496$ cities in Nicaragua. To obtain a $(PCP)$ instance, each city both constitutes a client and a candidate site. The number of sites to open is $p=5$.}

\begin{figure}[H]
\centering
\begin{subfigure}[t]{0.45\textwidth}
    \includegraphics[width=\textwidth]{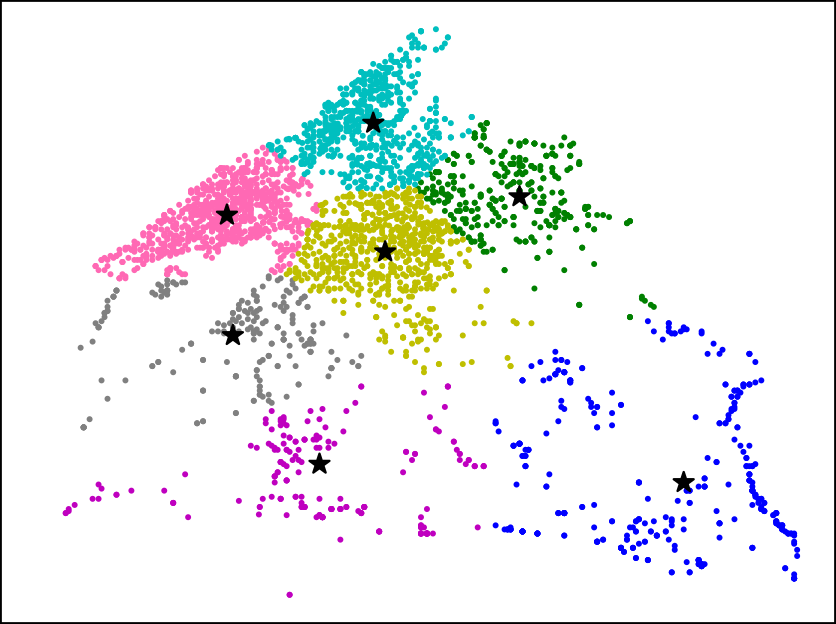}
    \caption{The 7 clusters and their medoids (initial representatives).}
    \label{fig:first}
\end{subfigure}
\hfill
\begin{subfigure}[t]{0.45\textwidth}
    \includegraphics[width=\textwidth]{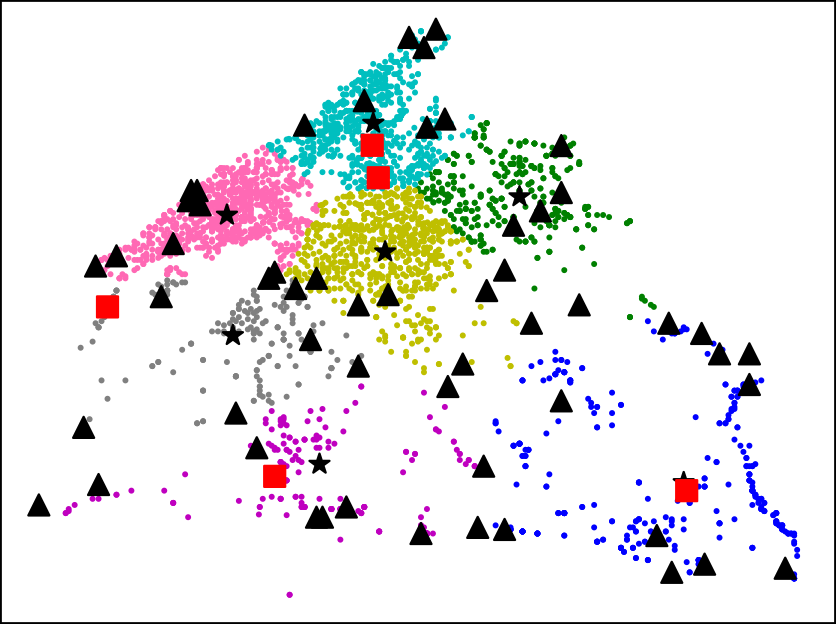}
    \caption{It. 1: optimal solution when distances are rounded down to multiples of $10^3$. {This iterations adds $55$ new representatives and provides the bounds $[1000, 1603]$.}}
    \label{fig:second}
\end{subfigure}
\vspace{.3cm}

\begin{subfigure}{0.45\textwidth}
    \includegraphics[width=\textwidth]{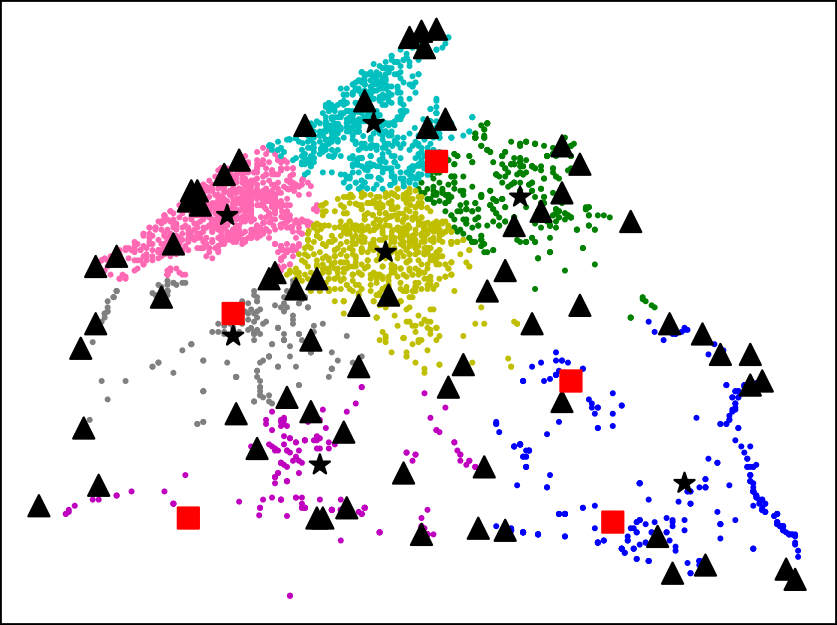}
    \caption{It. 2: optimal solution when distances are rounded down to multiples of $10^2$. {This iterations adds $13$ new representatives and provide the bounds $[1100, 1197]$.}}
    \label{fig:third}
\end{subfigure}
\hfill
\begin{subfigure}{0.45\textwidth}
    \includegraphics[width=\textwidth]{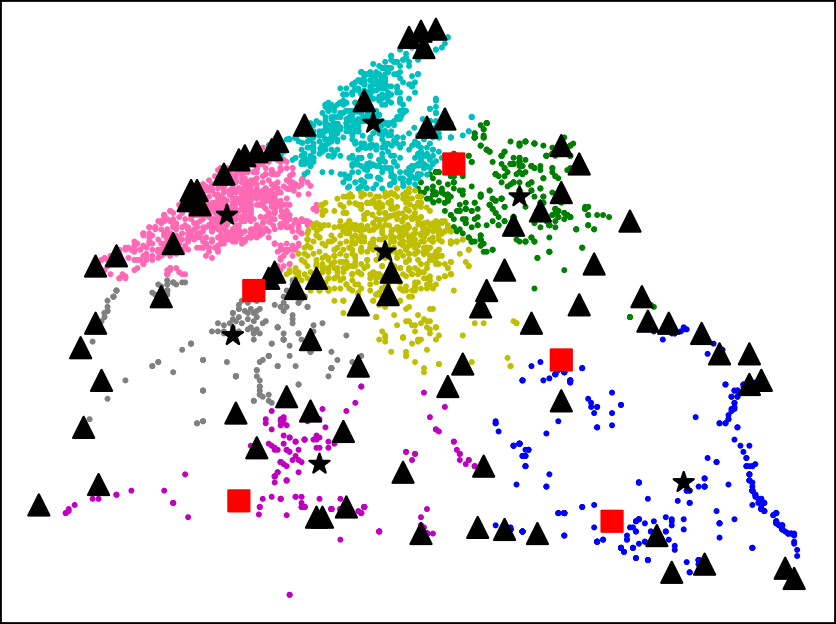}
    \caption{It. 3: optimal solution when distances are rounded down to multiples of $10^1$. {This iterations adds $11$ new representatives and provide the bounds $[1120, 1123]$.}}
    \label{fig:fourth}
\end{subfigure}        
\caption{Illustration of our algorithm on instance \texttt{nu3496} with $p=5$. {The medoids of the clusters are represented by stars, the other clients in the representative set by triangles and the sites opened in the optimal solution of each iteration by red squares.}}
\label{fig:illustration}
\end{figure}

{Figure~\ref{fig:illustration} represents the main steps of our algorithm. In Figure~\ref{fig:first}, the partitioning of the clients in $7$ clusters is depicted. Each cluster is represented by a distinct colour and by its star-shaped medoid. For this instance the parameter $\alpha$ is initially set to $3$. Consequently, in the first iteration we solve the problem $(PCP_{\alpha=3})$ in which all distances are rounded down to their closest multiple of  $1000$. During the resolution of $(PCP_{3})$, $55$ new clients are added to the set of representatives. These clients are represented by triangles in Figure~\ref{fig:second} and the red squares correspond to the opened sites in the optimal solution of $(PCP_3)$. The optimal radius is $1000$, which provides a lower bound on the radius of the original $(PCP)$ problem. Moreover, when considering the true distances (i.e., without rounding them down), this same  solution leads to a radius of $1603$ which thus constitutes an upper bound.}

{In the next iteration, we solve $(PCP_{\alpha=2})$ where the distances are rounded down to their closest multiple of $100$. Thanks to the bounds $[1000, 1603]$ obtained in the previous iteration, we know that there are only $7$ possible values for the optimal radius of $(PCP_2)$: $1000$, $1100$, $1200$, $1300$, $1400$, $1500$, or $1600$. More generally, in our algorithm, we necessarily solve $(PCP_\alpha)$ problems containing at most 10 distances. As represented in Figure~\ref{fig:third}, we only need to add $13$ additional clients to the set of representatives to compute the optimal solution of $(PCP_2)$. We deduce from this solution that the optimal radius of the original problem is in the interval $[1100, 1197]$.}

{The third iteration, represented in Figure~\ref{fig:fourth}, adds $11$ clients to the set of representatives to solve $(PCP_1)$ and ensures that the optimal radius is between $1120$ and $1123$. Finally, an optimal radius of $1123$ is obtained during the last iteration in which no new client is added to the set of representatives. In the end, our algorithm optimally solves the problem by only considering  $86$ clients out of the 3496 contained in the instance.}

\vspace{0.3cm}


\section{Computational experiments} \label{sec:experPCP}

Following previous works~\cite{Contardo2019,GAAR2022}, we consider in this experiment 44 instances from the TSPLib~\cite{TSPlib} containing the coordinates of $1600$ to $1904711$ cities. For each TSP instance, a $p$-center instance is obtained by considering a client and a potential site at the location of each city (i.e. $|\cliset|=|\siteset|$). The values of $p$ considered are $p\in \s{2, 3, 5, 10, 15, 20, 50, 100, 200}$. {Note that $p$ is at most $30$ in the results presented in~\cite{Contardo2019} and~\cite{GAAR2022}.} A time limit of $3$ hours is considered. For all instances, the distances between the clients and the sites are obtained by computing the euclidean distance between their coordinates rounded to its closest integer. For our experiments, we consider an Intel XEON E5-2643 processor 3,4 GHz, with 24 threads and 252 GB of RAM.

First, in Section~\ref{sec:expeFeatures} we evaluate the efficiency of three of our algorithmic features: the site dominations, the local search, and the iterative rounding of the distances. This enables to identify that the use of site dominations on the largest instance is too time consuming and should be deactivated. Then, in Section~\ref{sec:expeMethods} we compare our approach to the two best approaches from the litterature: the iterative algorithm from~\cite{Contardo2019} and the branch-and-cut from~\cite{GAAR2022}. We kindly thank the authors of these articles for providing us with the resources to execute their methods on our computer, thus allowing a fair comparison. All MILPs are solved using IBM ILOG CPLEX 22.1.

\subsection{Features comparison {of our method}}
\label{sec:expeFeatures}

We first compare the performances of our method in four different configurations:
\begin{enumerate}
    \item \textit{All features activated}: we use Algorithm~\ref{alg:modulo} to solve the $(PCP)$ which calls Algorithm~\ref{alg:clustering} at each iteration to solve $(PCP_\alpha)$.
    \item \textit{Without the local search}: we do not use the local search algorithm at step~\ref{s:alt} of Algorithm~\ref{alg:clustering}. Consequently, at step~\ref{s:quadrant} only solution $\hat y$ is considered to add clients to $\repset$ instead of $\s{\hat y, y^1, ..., y^S}$.
    \item \textit{Without dominations}: The non-dominated sites $J^\alpha_\repset$ are not computed in step~\ref{s:dom} of Algorithm~\ref{alg:clustering}, nor updated in step~\ref{s:dom} Thus, $(PCP_\alpha)$ is always solved by considering all the sites instead of only the non-dominated ones.
    \item \textit{Without rounding the distances}: we directly solve $(PCP_{\alpha=0})$ through Algorithm~\ref{alg:clustering} without using Algorithm~\ref{alg:modulo}. As a consequence, the number of distinct distances between clients and sites may be high, which is likely to increase significantly the number of set cover iterations.
\end{enumerate}

\begin{figure}[!ht]
    \centering
    \includegraphics[width=1\linewidth]{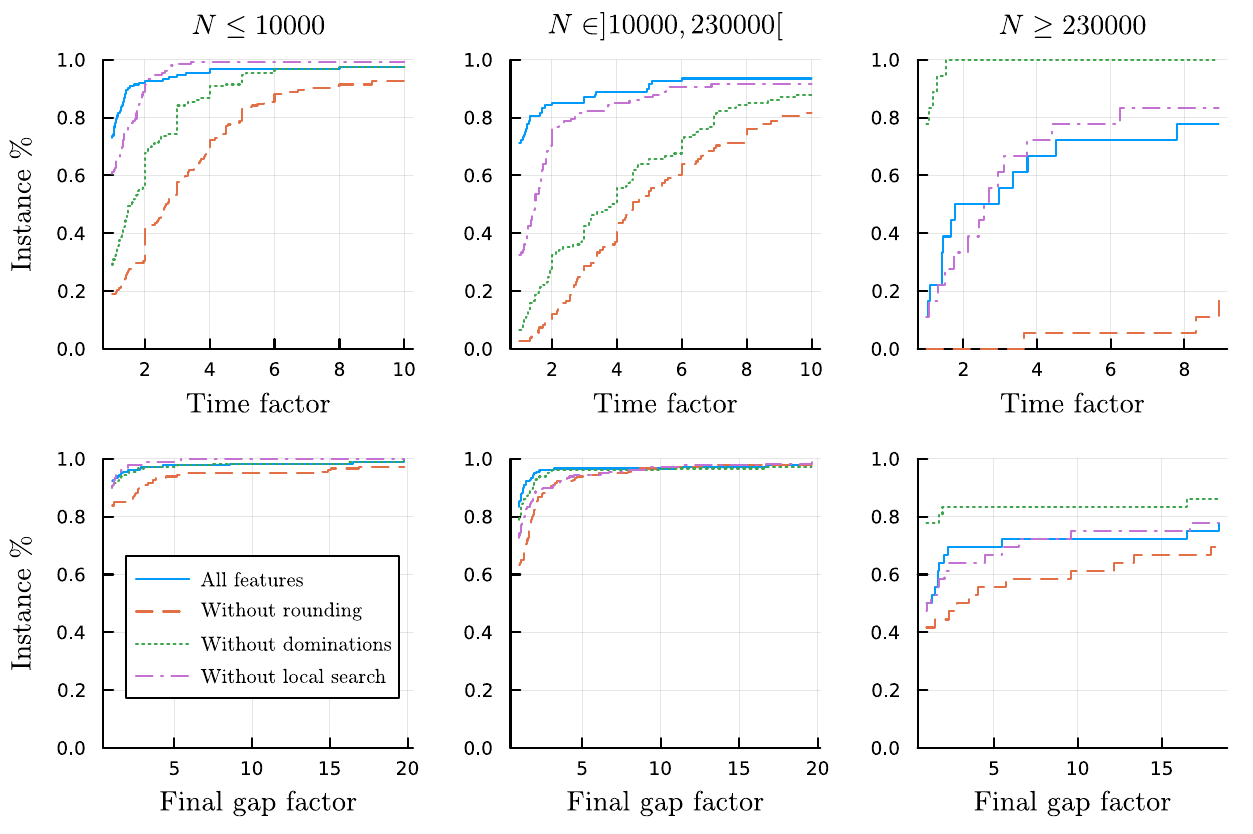}
    \caption{Performance profiles of our algorithm with different features deactivated.}
    \label{fig:featureResults}
\end{figure}

The results are summarized in Figure~\ref{fig:featureResults} through performance profiles of CPU times and gap. Each curve of the performance profile corresponds to the results of one configuration. Each point $(x, y)$ of a curve indicates that  for $x$ percents of the instances, the configuration was no more than $y$ times worse than the best method. In particular, $y=1$ corresponds to the proportion of instances on which the configuration is the best.

\newpage
The results in Figure~\ref{fig:featureResults} are split in three categories depending on the size of the instance: $N\leq 10000$, $N\in[10000, 230000]$, and $N\geq 230000$. For each size, we represent two performance profiles: one for the solution time, one for the final gap obtained. Note that for a given configuration, the resolution time of an instance is only accounted for if it is below the time limit.

We can first see in Figure~\ref{fig:featureResults} that the configuration in which the distance are not rounded is always the worst one, both in terms of gap and of resolution time. This is even more significant for the largest instances. The use of Algorithm~\ref{alg:modulo} which iteratively rounds down the distances more and more precisely, thus seems to be the most efficient feature of our algorithm. We can then see that the computation of the site dominations enables to speed up the resolution except for the largest instances. For these instances, most of the available time is spent computing the dominations instead of solving the set cover problems which eventually leads to poor upper and lower bounds. As a consequence, in the next section we do not compute the dominations for these instances. Finally, we can see that removing the local search tends to slow down the resolution but enables to improve the number of instances solved to optimality. Since our objective is to improve the solution of hard instances, we keep this feature in the following.


\subsection{Comparison to state-of-the-art methods}
\label{sec:expeMethods}

In this section, we compare our algorithm with the iterative algorithm from~\cite{Contardo2019} and the branch-and-cut from~\cite{GAAR2022}. As mentioned before, thanks to their authors, we were able to run the three methods on the same computer and execute their code for larger values of $p$ with the same time limit than us.

\begin{figure}[htpb]
    \centering
    \includegraphics[width=1\linewidth]{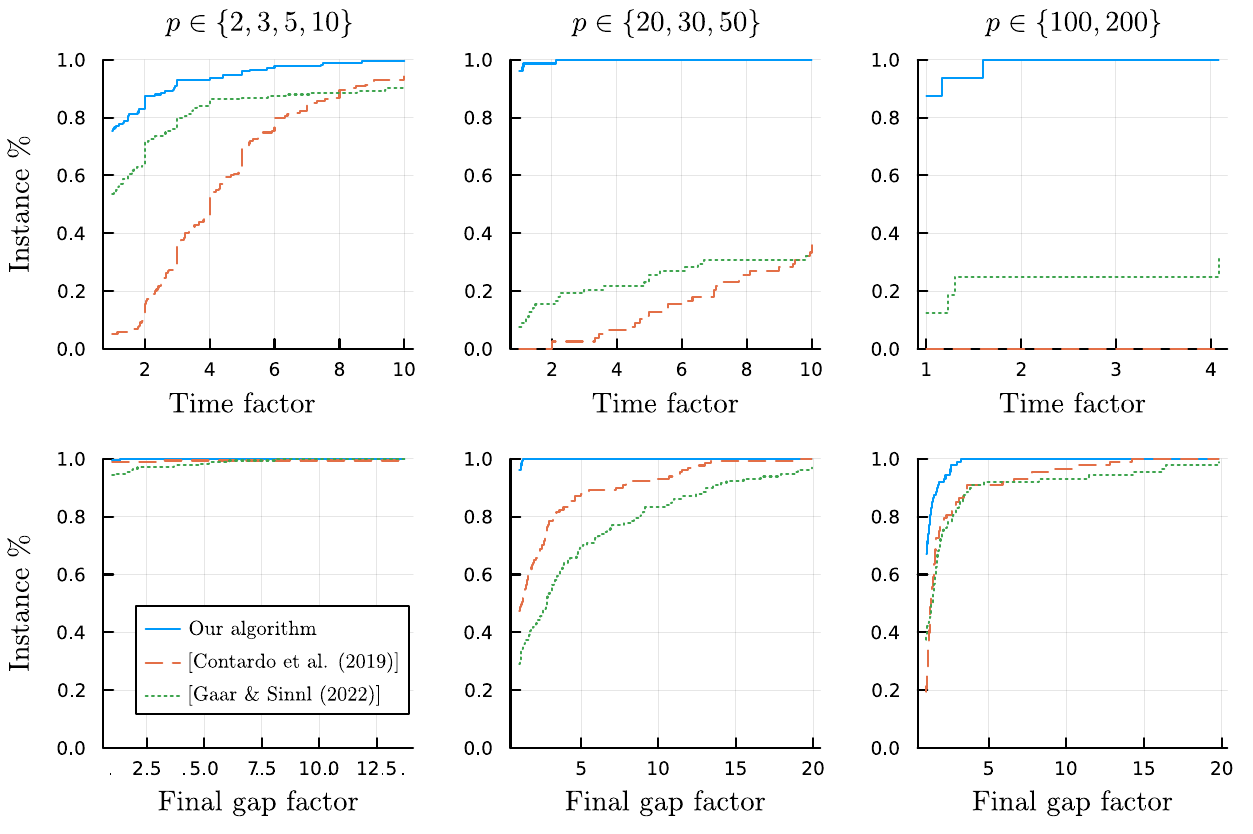}
    \caption{Performance profiles of our algorithm compared with two state-of-the-art methods.}
    \label{fig:methodResults}
\end{figure}

The results are summarized by the performance profiles in Figure~\ref{fig:methodResults} which splits the instances in three categories depending on the number of sites opened: $p\in\{2, 3, 5, 10\}$, $p\in\{20, 30, 50\}$, and $p\in\{100, 200\}$. To give a more detailed view of the results, we also present Tables~\ref{tab:p} and~\ref{tab:instance} in which the results are averaged for each value of $p$ and $N$, respectively. In these tables, a value is in bold if it is strictly better than the one obtained by the two other methods. 

\newpage
From Table~\ref{tab:p}, we can see that the three methods solve all instances to optimality within $3$ hours for the smallest values of $p\in\s{2, 3, 5}$. For these instances, the branch-and-cut from~\cite{GAAR2022} is the fastest. However, for all the other values of $p$, our algorithm provides the best results in terms of number of instances solved to optimality, final gap and resolution time. In particular, the resolution time reduction when applying our method for $p\geq 10$ is very significant: from $3$ times faster for $p=10$ to $60$ times faster for $p=200$. This is clearly highlighted by the two top left profiles of Figure~\ref{fig:methodResults}.

  \begin{table}[htpb]
    \centering
    \resizebox{\textwidth}{!}{%
    \begin{tabular}{|@{~}c@{~}c|c@{~}r@{~}r|c@{~}r@{~}r|c@{~}r@{~}r@{~}|}
    \hline
        \multirow{2}{*}{$p$} & \multirow{2}{*}{\#inst.}& \multicolumn{3}{c|}{Our algorithm} & \multicolumn{3}{c|}{\cite{Contardo2019}} &  \multicolumn{3}{c|}{\cite{GAAR2022}} \\ 
      & & \#opt. &  gap[\%] &  time[s]$^*$ & \#opt. &  gap[\%] &  time[s]$^*$ & \#opt. &  gap[\%] &  time[s]$^*$ \\ \hline
        2 & 44 & 44 & 0 & 63 & 44 & 0 & 22 & 44 & 0 & \textbf{9} \\ 
        3 & 44 & 44 & 0 & 74 & 44 & 0 & 53 & 44 & 0 & \textbf{22} \\ 
        5 & 44 & 44 & 0 & 144 & 44 & 0 & 252 & 44 & 0 & \textbf{124} \\ 
        10 & 44 &\textbf{43} &\textbf{0.3} & \textbf{153} & 41 & 0.6 & 465 & 33 & 1.2 & 1098 \\ 
        15 & 44 &\textbf{35} &\textbf{0.9} & \textbf{21} & 30 & 2.5 & 182 & 19 & 6 & 509 \\ 
        20 & 44 &\textbf{30} &\textbf{1.9} & \textbf{107} & 20 & 5.7 & 542 & 13 & 10.9 & 1384 \\ 
        50 & 44 &\textbf{13} &\textbf{10.7} & \textbf{19} & 7 & 16.6 & 812 & 5 & 22 & 1251 \\ 
        100 & 44 &\textbf{9} &\textbf{24.4} & \textbf{19} & 6 & 25.6 & 660 & 3 & 25.8 & 677 \\ 
        200 & 44 &\textbf{7} &\textbf{27.4} & \textbf{18} & 5 & 32.4 & 1094 & 4 & 31.5 & 1779 \\ \hline
        sum/avg.	&396&	\textbf{269}&	\textbf{7.3}	&\textbf{94} &	241	&9.2&	242&	209&	10.8&	406\\
        \hline

    \end{tabular}
    }
    \vspace{.1cm}
    
    $^*$: only computed on instances optimally solved by the three methods.\hfill~

    \caption{Average results obtained for each method and each value of parameter $p$. A value is in \textbf{bold} if it is strictly better than that of the other methods. All methods run on the same computer with a time limit of 3 hours.}
    \label{tab:p}
\end{table}

In Table~\ref{tab:instance}, the bold values confirm that for most of the instances, we outperform the two state-of-the-art methods. The method from~\cite{Contardo2019} is able to obtain better gaps on $9$ instances and the branch-and-cut from~\cite{GAAR2022} is faster than our method on $5$ instances. Four of these five instances are the largest ones. This coincides with the observation in Section~\ref{sec:expeFeatures} that computing the domination for these instances was too time consuming. Nevertheless, the number of instances solved to optimality and the final gaps obtained by our method remains better in average for these larges instances than that of the branch-and-cut.

\begin{table}[htpb]
\begin{small}
    \Centering
    \resizebox{0.98\textwidth}{!}{%
    \begin{tabular}{|@{~}c@{~}c@{~}c|c@{~}r@{~}r|c@{~}r@{~}r|c@{~}r@{~}r@{~}|}
    \hline
        \multirow{2}{*}{Name} & \multirow{2}{*}{$N$} & \multirow{2}{*}{\#inst.}& \multicolumn{3}{c|}{Our algorithm}  & \multicolumn{3}{c|}{\cite{Contardo2019}}
    & \multicolumn{3}{c|}{\cite{GAAR2022}}  \\ 
         &&  & \#opt. &  gap[\%] &  time[s]$^*$ & \#opt. &  gap[\%] &  time[s]$^*$ & \#opt. &  gap[\%] &  time[s]$^*$ \\ \hline
        rw&1621 & 9 & 9 & 0 & \textbf{3} & 9 & 0 & 49 & 9 & 0 & 14 \\ 
        u&1817 & 9 &\textbf{9} &\textbf{0} & \textbf{25} & 8 & 1.2 & 770 & 8 & 0.2 & 1694 \\ 
        rl&1889 & 9 &\textbf{9} &\textbf{0} & \textbf{7} & 8 & 0.3 & 142 & 6 & 0.6 & 1045 \\ 
        mu&1979 & 9 & 9 & 0 & \textbf{4} & 9 & 0 & 122 & 9 & 0 & 227 \\ 
        pr&2392 & 9 &\textbf{8} &\textbf{0.6} & \textbf{4} & 6 & 4.1 & 25 & 4 & 4.5 & 196 \\ 
        pcb&3038 & 9 &\textbf{6} &\textbf{8.2} & \textbf{4} & 5 & 8.3 & 20 & 4 & 9.5 & 31 \\ 
        nu&3496 & 9 &\textbf{9} &\textbf{0} & \textbf{7} & 7 & 2.2 & 103 & 7 & 2.7 & 778 \\ 
        ca&4663 & 9 &\textbf{9} &\textbf{0} & \textbf{1} & 8 & 1 & 14 & 6 & 1.9 & 4 \\ 
        rl&5915 & 9 &\textbf{6} &\textbf{6.6} & \textbf{1} & 5 & 9 & 5 & 4 & 12.2 & 3 \\ 
        rl&5934 & 9 & 6 &\textbf{5.6} & \textbf{17} & 6 & 7.5 & 360 & 4 & 9.5 & 658 \\ 
        tz&6117 & 9 &\textbf{7} &\textbf{1.5} & \textbf{6} & 6 & 6.1 & 31 & 4 & 8.9 & 19 \\ 
        eg&7146 & 9 & 8 &\textbf{0.3} & 4 & 8 & 2.4 & 28 & 7 & 5.3 & 4 \\ 
        pla&7397 & 9 &\textbf{7} &\textbf{2.2} & \textbf{4} & 6 & 5.5 & 51 & 5 & 8.8 & 35 \\ 
        ym&7663 & 9 &\textbf{7} & 6.5 & \textbf{4} & 6 & \textbf{4.3} & 31 & 6 & 8 & 12 \\ 
        pm&8079 & 9 &\textbf{8} &\textbf{0.8} & \textbf{10} & 7 & 3 & 478 & 7 & 5.3 & 152 \\ 
        ei&8246 & 9 &\textbf{6} & 10.7 & \textbf{7} & 5 & \textbf{9.7} & 46 & 4 & 10.6 & 30 \\ 
        ar&9152 & 9 & 6 &\textbf{2.7} & \textbf{210} & 6 & 8.8 & 691 & 6 & 8.6 & 2377 \\ 
        ja&9847 & 9 & 7 &\textbf{3.7} & \textbf{3} & 7 & 4.1 & 16 & 6 & 8 & 4 \\ 
        gr&9882 & 9 &\textbf{7} & 7.1 & \textbf{6} & 6 & \textbf{4.6} & 62 & 6 & 8.7 & 23 \\ 
        kz&9976 & 9 & 6 &\textbf{2.1} & \textbf{11} & 6 & 7.6 & 96 & 5 & 10.6 & 31 \\ 
        fi&10639 & 9 &\textbf{6} &\textbf{5.3} & 9 & 5 & 10.6 & 21 & 4 & 10.7 & \textbf{8} \\ 
        rl&11849 & 9 &\textbf{5} &\textbf{6.1} & \textbf{17} & 4 & 10.6 & 66 & 4 & 12 & 767 \\ 
        usa&13509 & 9 &\textbf{6} &\textbf{6.8} & \textbf{8} & 5 & 11.6 & 87 & 4 & 12.1 & 158 \\ 
        brd&14051 & 9 &\textbf{6} & 11.5 & \textbf{14} & 5 & \textbf{10.1} & 39 & 4 & 11.7 & 2353 \\ 
        mo&14185 & 9 & 6 &\textbf{4.9} & \textbf{6} & 6 & 8 & 38 & 4 & 11.8 & 14 \\ 
        ho&14473 & 9 & 6 & 8.6 & \textbf{6} & 6 & \textbf{7.2} & 23 & 4 & 11.3 & 29 \\ 
        d&15112 & 9 &\textbf{5} & 18.4 & \textbf{7} & 4 & \textbf{11.8} & 20 & 4 & 14 & 18 \\ 
        it&16862 & 9 & 6 &\textbf{5.6} & \textbf{14} & 6 & 7.2 & 106 & 6 & 10.1 & 50 \\ 
        d&18512 & 9 &\textbf{5} &\textbf{8.8} & \textbf{7} & 4 & 12.6 & 21 & 3 & 13.1 & 17 \\ 
        vm&22775 & 9 & 6 &\textbf{5.8} & \textbf{8} & 6 & 8 & 61 & 5 & 10.8 & 12 \\ 
        sw&24978 & 9 &\textbf{6} &\textbf{5.7} & \textbf{13} & 5 & 10.5 & 36 & 4 & 16.4 & 22 \\ 
        fyg&28534 & 9 & 4 &\textbf{11.8} & \textbf{100} & 4 & 13.6 & 867 & 4 & 15.8 & 1156 \\ 
        bm&33708 & 9 &\textbf{6} &\textbf{6.2} & \textbf{32} & 5 & 10 & 134 & 4 & 12.2 & 238 \\ 
        pla&33810 & 9 & 4 &\textbf{9.3} & \textbf{25} & 4 & 15.3 & 283 & 3 & 16.1 & 70 \\ 
        bby&34656 & 9 & 4 &\textbf{11.5} & \textbf{130} & 4 & 12.8 & 737 & 4 & 15.4 & 484 \\ 
        pba&38478 & 9 & 4 &\textbf{12.4} & \textbf{16} & 4 & 13.5 & 51 & 3 & 16.5 & 24 \\ 
 ch&71009&	9&	5&	14.5&	\textbf{94} &	5&	\textbf{12.9}&	327&	4&	14.2&	509\\
pla&85900	&9	&\textbf{4} &	\textbf{9.3} &	\textbf{76}&	3&	18.9&	840	&3&	18&	306\\
       sra&104815 & 9 &\textbf{5} & 16.4 & 61 & 4 & \textbf{12.8} & 96 & 3 & 17.1 & \textbf{46} \\ 
        usa&115475 & 9 & 4 &\textbf{13.9} & \textbf{240} & 4 & 15.8 & 1048 & 4 & 15.8 & 1678 \\ 
        ara&238025 & 9 & 4 & 19.5 & 288 & 4 & \textbf{16.1} & 230 & 3 & 16.3 & \textbf{124} \\ 
        lra&498378 & 9 &\textbf{4} &\textbf{12.5} & 579 & 3 & 19.4 & 922 & 3 & 12.6 & \textbf{230} \\ 
        lrb&744710 & 9 & 3 & 18.2 & 990 & 3 & 22.3 & 1370 & 3 & 18.2 & \textbf{984} \\ 
        world&1904711 & 9 &\textbf{6} &\textbf{19} & 2076 & 4 & 25.3 & 812 & 5 & 30.3 & \textbf{272} \\ \hline
&        sum/avg.	&  396	& \textbf{269}&	\textbf{7.3}	&\textbf{94}&	241	&9.2 &	242 &	209	& 10.8 &	406\\\hline
    \end{tabular}
    }
    \end{small}
    \vspace{.1cm}
    
    $^*$: only computed on instances optimally solved by the three methods.\hfill~

    \caption{Average results obtained for each method and each instance size. A value is in \textbf{bold} if it is strictly better than that of the other methods. All methods run on the same computer with a time limit of 3 hours.}
    \label{tab:instance}
  \end{table}

{ }
\newpage
\section{Conclusions} \label{sec:conclusionPCP}

We introduce a new exact method to solve the $p$-center problem based on an iterative rounding of the distances which greatly limits the number of possible value for the optimal solution at each iteration.  {Our experimental analysis highlights that this new feature is by far the one that most improves the performances of our algorithm and allows it to scale up. Moreover, our method}  includes a clustering of the clients which enables to efficiently initialize the subset of representatives~$\repset$. {We  also introduce the notion of cluster quadrants which enables to update~$\repset$ by adding to it relevant clients which are not close to one another and not too many either.} Finally, we also efficiently manage the computation of the site dominations and limit the number of MILP subproblems to solve by considering their linear relaxation as often as possible. Through extensive experiments, we compare our approaches to the iterative algorithm from~\cite{Contardo2019} and the branch-and-cut from~\cite{GAAR2022}. As soon as parameter $p$ is greater than $5$, our method significantly outperforms these two state-of-the-art algorithms in terms of resolution time, final gap and number of instances solved to optimality. In the future we will try to better handle site dominations of the largest instances and to speed up the update of the subset of clients.




  \section*{Acknowledgments}
    

     \noindent  We would like to thank the authors of~\cite{GAAR2022} and \cite{Contardo2019} for providing us with the executable and source code of their methods, respectively.

\bibliographystyle{apalike}
    \bibliography{references}

    \end{document}